\documentclass[aps,prd,preprint,nofootinbib,superscriptaddress]{revtex4-1}
\usepackage{amsmath}
\usepackage{latexsym}
\usepackage{amsfonts}
\usepackage{graphicx}
\usepackage{mathrsfs}
\usepackage{epstopdf}
\usepackage{subfigure}
\begin{document}

\title{On the constant-roll inflation with large and small $\eta_H$}

\author{Qing Gao}
\email{gaoqing1024@swu.edu.cn}
\affiliation{School of Physical Science and Technology, Southwest University, Chongqing 400715, China}

\author{Yungui Gong}
\email{yggong@mail.hust.edu.cn}
\affiliation{School of Physics, Huazhong University of Science and Technology,
Wuhan, Hubei 430074, China}

\author{Zhu Yi}
\email{yizhu92@hust.edu.cn}
\affiliation{School of Physics, Huazhong University of Science and Technology,
Wuhan, Hubei 430074, China}

\begin{abstract}
We study the seemingly duality between large and small $\eta_H$ for the constant-roll inflation with the second slow-roll parameter $\eta_H$ being a constant.
In the previous studies, only the constant-roll inflationary models with small $\eta_H$ are found to be consistent with the observations.
The seemingly duality suggests that the constant-roll inflationary models with large $\eta_H$ may be also consistent with the observations.
We find that the duality between the constant-roll inflation with large and small $\eta_H$ does not exist
because both the background and scalar perturbation evolutions are very different.
By fitting the constant-roll inflationary models to the observations,
we get $-0.016\le\eta_H\le-0.0078$ at the 95\% C.L if we take $N=60$ for the models with increasing $\epsilon_H$ in which inflation ends when $\epsilon_H=1$,
and $3.0135\le \eta_H\le 3.021$ at the 68\% C.L., and $3.0115\le \eta_H\le 3.024$ at the 95\% C.L. for the models with decreasing $\epsilon_H$.
\end{abstract}


\maketitle


\section{Introduction}\label{sec1}

Inflation explains the flatness and horizon problems in standard cosmology,
and the quantum fluctuations of the inflaton seed the large scale structure of the Universe and
leave imprints on the cosmic microwave background radiation \cite{Guth:1980zm,Linde:1981mu,Albrecht:1982wi,Starobinsky:1980te,Guth:1982ec}.
To solve the problems such as the flatness, horizon and monopole problems,
the number of $e$-folds remaining before the end of inflation must be large enough
and it is usually taken to be $N=50-60$ due to the uncertainties in reheating physics.
This requires the potential of the inflaton to be nearly flat, i.e., the slow-roll inflation.
The temperature and polarization
measurements on the cosmic microwave background anisotropy
conformed the nearly scale invariant power spectra predicted by the slow-roll inflation and gave the constraints $n_s=0.965\pm 0.004$ (68\% C.L.)
and $r_{0.05}<0.06$ (95\% C.L.) \cite{Akrami:2018odb,Ade:2018gkx}.

Recently, the constant-roll inflation with $\eta_H$ being a constant \cite{Martin:2012pe,Motohashi:2014ppa} attracted some attentions
because the inflationary potential and the background equation of motion can be solved analytically.
The slow-roll parameter $\eta_H$ is a constant and it may not be small, the model is different from the typical slow-roll inflationary models.
In particular, when the inflationary potential becomes very flat, $\eta_H=3$, we get the ultra slow-roll inflation \cite{Tsamis:2003px,Kinney:2005vj}.
Due to the violation of the slow-roll condition, the curvature perturbation may evolve outside the horizon and the slow-roll results may not be applied \cite{Leach:2000yw,Leach:2001zf,Kinney:2005vj,Jain:2007au,Namjoo:2012aa,Martin:2012pe,Motohashi:2014ppa,Yi:2017mxs}.
However, for the constant-roll inflation with $\eta_H>1$, the slow-roll parameter $\epsilon_H$ decreases with time and is small during inflation,
so we can still use the standard method of Bessel function approximation to calculate the power spectra.
Neglecting the contribution from $\epsilon_H$, it was found there exists a duality between the ultra slow-roll inflation and the slow-roll inflation \cite{Tzirakis:2007bf,Morse:2018kda},
i.e., if we replace $\eta_H$ by $\bar{\eta}_H=3-\eta_H$, we get the same result for the scalar spectral tilt.
Recall that the observational data constrained $\eta_H$ to be small \cite{Motohashi:2017aob,Gao:2018cpp,GalvezGhersi:2018haa}, these results are in conflict with
the duality relation, so it is necessary to revisit the observational constraint to include the constraint on the ultra-slow inflation.
For the ultra slow-roll inflation, it is legitimate to neglect $\epsilon_H$. For the typical slow-roll inflation, $\epsilon_H$ and $\eta_H$ are in the same order,
so $\epsilon_H$ cannot be neglected and it is interesting to discuss the duality up to the first order of $\epsilon_H$ in the constant-roll inflation.
The difference in $\epsilon_H$ may cause different amplitudes for the power spectra or different energy scale of inflation.
Furthermore, due to the smallness of $\epsilon_H$ in the ultra slow-roll inflation, it can be used to generate a large curvature perturbation at small scales which produces
primordial black holes and secondary gravitational waves \cite{Germani:2017bcs,Motohashi:2017kbs,Gong:2017qlj}.
For more discussion on the constant-roll inflation, please see Refs. \cite{Motohashi:2017vdc,Oikonomou:2017bjx,Odintsov:2017qpp,Nojiri:2017qvx,
Dimopoulos:2017ged,Ito:2017bnn,Karam:2017rpw,Fei:2017fub,Cicciarella:2017nls,Anguelova:2017djf,Gao:2018tdb,Mohammadi:2018wfk,Pattison:2018bct}.

In this paper, we extend the discussion of the duality between the ultra slow-roll inflation and the slow-roll inflation to include the effect of $\epsilon_H$.
The paper is organized as follows.
In the Sec. \ref{sec2}, we review the constant-roll inflation and discuss the duality between the ultra slow-roll inflation with large constant $\eta_H$ and
the slow-roll inflation with small constant $\eta_H$.
In Sec. \ref{sec3}, we fit constant roll models to the observational data. The conclusions are drawn in Sec. \ref{sec5}.

\section{The constant-roll inflation}
\label{sec2}

We use the Hubble flow slow-roll parameters \cite{Liddle:1994dx},
\begin{equation}
\label{hfslr1}
^n\beta_H=2\left(\frac{(H_{,\phi})^{n-1}H^{(n+1)}_{,\phi}}{H^n}\right)^{1/n},
\end{equation}
where $H_{,\phi}=dH/d\phi$ and $H^{(n)}_{,\phi}=d^nH/d\phi^n$. In particular, the first three slow-roll parameters are
\begin{equation}
\label{hfslr2}
\epsilon_H=2\left(\frac{H_{,\phi}}{H}\right)^2=-\frac{\dot H}{H^2},
\end{equation}
\begin{equation}
\label{hfslr3}
\eta_H=\frac{2 H_{,\phi}^{(2)}}{H}=-\frac{\ddot\phi}{H\dot\phi}=-\frac{\ddot H}{2H\dot{H}},
\end{equation}
\begin{equation}
\label{xih}
\xi_H=\frac{4H_{,\phi}H_{,\phi}^{(3)}}{H^2}=\frac{\dddot H}{2H^2\dot H}-2\eta_H^2,
\end{equation}
and the evolution of the slow-roll parameters are
\begin{equation}
\label{dotep}
\dot{\epsilon}_H=2H\epsilon_H(\epsilon_H-\eta_H),
\end{equation}
\begin{equation}
\label{doteta}
\dot{\eta}_H=H(\epsilon_H\eta_H-\xi_H),
\end{equation}
where $\dot{H}=dH/dt$. For the constant-roll inflation with constant $\eta_H$, we get $\xi_H=\epsilon_H\eta_H$.
From Eq. \eqref{dotep}, we see that if $\epsilon_H>\eta_H$, then $\epsilon_H$ increases monotonically with time.
Otherwise, if $\epsilon_H<\eta_H$, then $\epsilon_H$ decreases monotonically with time.
Since $\epsilon_H\le 1$, so $\epsilon_H$ decreases monotonically with time for the constant-roll inflationary model with $\eta_H>1$, such as the ultra slow-roll inflation with $\eta_H\approx 3$.

The scalar perturbation is governed by Mukhanov-Sasaki equation \cite{Mukhanov:1985rz,Sasaki:1986hm},
\begin{equation}
\label{eq21}
v_k'' + \left(k^2 - \frac{z''}{z} \right)v_k = 0,
\end{equation}
where
\begin{equation}
\label{normvars}
z = \frac{a\dot \phi}{H},
\end{equation}
$v_k'=dv_k/d\tau$, $\tau$ is the conformal time, and the mode function $v_k$ for a Fourier mode is related with the curvature perturbation $\zeta$ by $v_k=z \zeta_k$.
To the first order of $\epsilon_H$, $aH \approx -(1+\epsilon_H)/\tau$,
and Eq. \eqref{eq21} becomes
\begin{equation}
\label{eq21a}
v_k'' + \left(k^2 - \frac{\nu^2-1/4}{\tau^2} \right)v_k = 0,
\end{equation}
where
\begin{equation}
\label{slreq9}
\nu\approx\frac{1}{2}|2\eta_H-3|+\frac{(2\eta_H^2-9\eta_H+6)\epsilon_H}
{|2\eta_H-3|}.
\end{equation}
Since $\eta_H$ is a constant
and the change of $\epsilon_H$ can be neglected which is true for both slow-roll and ultra slow-roll inflation
\footnote{For the ultra slow-roll inflation, $\epsilon_H$ can be very small because it decreases with time.}, so $\nu$ can be approximated as a constant,
the solution to Eq. \eqref{eq21a} for the mode function $v_k$ is the Hankel function of order $\nu$,
\begin{equation}
\label{modesol1}
v_k=\frac{\sqrt{\pi}}{2}e^{i(\nu+1/2)\pi/2} \sqrt{-\tau} H_\nu^{(1)}(-k\tau).
\end{equation}
Therefore, the power spectrum of the scalar perturbation is
\begin{equation}
\label{pkeq1}
P_{\zeta}=\frac{k^3}{2\pi^2}|\zeta_k|^2= \frac{2^{2\nu-3}}{2\epsilon_H}\left[\frac{\Gamma(\nu)}{\Gamma(3/2)}\right]^2
\left(1+\epsilon_H\right)^{1-2\nu}\left(\frac{H}{2\pi}\right)^2\left(\frac{k}{aH}\right)^{3-2\nu}.
\end{equation}
The amplitude of the power spectrum at the horizon crossing is
\begin{equation}
\label{as}
A_s= \frac{2^{2\nu-3}}{2\epsilon_H}\left[\frac{\Gamma(\nu)}{\Gamma(3/2)}\right]^2
\left(1+\epsilon_H\right)^{1-2\nu}\left(\frac{H}{2\pi}\right)^2.
\end{equation}
The scalar spectral tilt is
\begin{equation}
\label{nseq1}
n_s-1= \frac{d\ln P_\zeta }{d\ln k} \approx 3-|2\eta_H-3|-\frac{2(2\eta_H^2-9\eta_H+6)\epsilon_H}
{|2\eta_H-3|}.
\end{equation}
Following the same procedure, we get the power spectrum of the tensor perturbation and the tensor to scalar ratio
\begin{equation}
\label{ra}
r\approx2^{3-|2\eta_H-3|}\left(\frac{\Gamma[3/2]}{\Gamma[|2\eta_H-3|/2]}\right)^216\epsilon_H.
\end{equation}
If we neglect the contribution of $\epsilon_H$ in Eqs. \eqref{slreq9}, \eqref{nseq1} and \eqref{ra},
we see that these expressions are unchanged if we replace $\eta_H$ by $\bar{\eta}_H=3-\eta_H$, i.e., there
exists a duality between $\eta_H$ and $\bar{\eta}_H=3-\eta_H$ as observed in \cite{Tzirakis:2007bf,Morse:2018kda}.
It this duality is true, then we can apply the usual slow-roll results to ultra slow-roll inflationary models.
In the previous analysis of the observational constraints on constant-roll inflation, only the model with small $\eta_H$ was found to
be consistent with the observations \cite{Motohashi:2017aob,Yi:2017mxs,Gao:2018cpp,GalvezGhersi:2018haa}.
This duality relation suggests that the ultra slow-roll inflationary models may also be consistent with the observations.
To investigate whether this is true, we discuss the issue of duality below.

\subsection{The constant-roll models}
From Eq. \eqref{hfslr3}, we get
\begin{equation}
\label{hsoleq1}
H(\phi)=c_1 \exp\left(\sqrt{\frac{\eta_H}{2}}\,\phi\right)+c_2 \exp\left(-\sqrt{\frac{\eta_H}{2}}\,\phi\right),
\end{equation}
for $\eta_H>0$. For $\eta_H<0$, the general solution is the form of trigonometric functions $\sin(x)$ and $\cos(x)$. Following Ref. \cite{Motohashi:2014ppa}, for $\eta_H>0$ we consider
the particular solutions
\begin{equation}
\label{hsoleq2}
H(\phi)=M\cosh\left(\sqrt{\frac{\eta_H}{2}}\,\phi\right),
\end{equation}
with the potential $V=3H^2-2(H_{,\phi})^2$,
\begin{equation}
\label{vsoleq1}
V(\phi)=M^2\left[3\cosh^2\left(\sqrt{\frac{\eta_H}{2}}\,\phi\right)-\eta_H\sinh^2\left(\sqrt{\frac{\eta_H}{2}}\,\phi\right)\right],
\end{equation}
and
\begin{gather}
\label{hsoleq3}
H(\phi)=M\sinh\left(\sqrt{\frac{\eta_H}{2}}\,\phi\right),\\
\label{vsoleq2}
V(\phi)=M^2\left[3\sinh^2\left(\sqrt{\frac{\eta_H}{2}}\,\phi\right)-\eta_H\cosh^2\left(\sqrt{\frac{\eta_H}{2}}\,\phi\right)\right].
\end{gather}
For $\eta_H<0$, the particular solutions are
\begin{gather}
\label{hsoleq4}
H(\phi)=M\cos\left(\sqrt{\frac{-\eta_H}{2}}\,\phi\right), \\
\label{vsoleq3}
V(\phi)=M^2\left[3\cos^2\left(\sqrt{\frac{-\eta_H}{2}}\,\phi\right)+\eta_H\sin^2\left(\sqrt{\frac{-\eta_H}{2}}\,\phi\right)\right],
\end{gather}
and
\begin{gather}
\label{hsoleq5}
H(\phi)=M\sin\left(\sqrt{\frac{-\eta_H}{2}}\,\phi\right),\\
\label{vsoleq4}
V(\phi)=M^2\left[3\sin^2\left(\sqrt{\frac{-\eta_H}{2}}\,\phi\right)+\eta_H\cos^2\left(\sqrt{\frac{-\eta_H}{2}}\,\phi\right)\right].
\end{gather}
For the constant-roll inflation, $H(\phi)$ is known, so $\dot\phi$ is determined from the relation $\dot\phi=-2H_{,\phi}$
\footnote{For these potentials, solutions other than the constant-roll inflation exist.}.
We don't consider the exponential solution because the corresponding power-law inflation is excluded by the observations.
The models \eqref{vsoleq1} and \eqref{vsoleq3} were studied in Refs. \cite{Motohashi:2014ppa,Motohashi:2017aob,Morse:2018kda,GalvezGhersi:2018haa}.
For the model \eqref{vsoleq1}, $\dot\epsilon_H<0$, so we need to introduce some mechanism to end inflation.
The model \eqref{vsoleq2} was studied in Ref. \cite{Yi:2017mxs}.
As discussed in Ref. \cite{Yi:2017mxs}, in the model \eqref{vsoleq2}, $\dot\epsilon_H>0$ and $\epsilon_H>\eta_H$,
so there is no inflation in this model if $\eta_H>1$, i.e., the model cannot support ultra slow-roll inflation and it is not applicable to the discussion of the duality relation.

\subsection{The duality between the slow-roll and the ultra slow-roll inflation}

For the slow-roll inflation with $\eta_H=\alpha$ and $|\alpha|\ll 1$, we get
\begin{equation}
\label{sraseq1}
A_s=\frac{1}{2\epsilon_H}\left(\frac{H}{2\pi}\right)^2,
\end{equation}
\begin{equation}
\label{nseqeta}
n_s-1=2\alpha -4\epsilon_H,
\end{equation}
and
\begin{equation}
\label{srreq1}
r=16\epsilon_H.
\end{equation}

For the ultra slow-roll inflation with $\eta_H=3-\alpha$ and $|\alpha|\ll 1$, we get
\begin{equation}
\label{usraseq1}
A_s=\frac{1}{2\epsilon_H}\left(\frac{H}{2\pi}\right)^2,
\end{equation}
\begin{equation}
\label{nseqeta1}
n_s-1= 2\alpha+2\epsilon_H,
\end{equation}
and
\begin{equation}
\label{usrreq1}
r=16\epsilon_H.
\end{equation}
From Eq. \eqref{nseqeta1}, we see that to be consistent with the observations $n_s<1$, we must take $\alpha<0$ because $\epsilon_H>0$,
so the constant-roll inflation with $\eta_H\gtrsim 3$ may be consistent with the observations.

\begin{figure}[htp]
\centering
\includegraphics[width=0.6\textwidth]{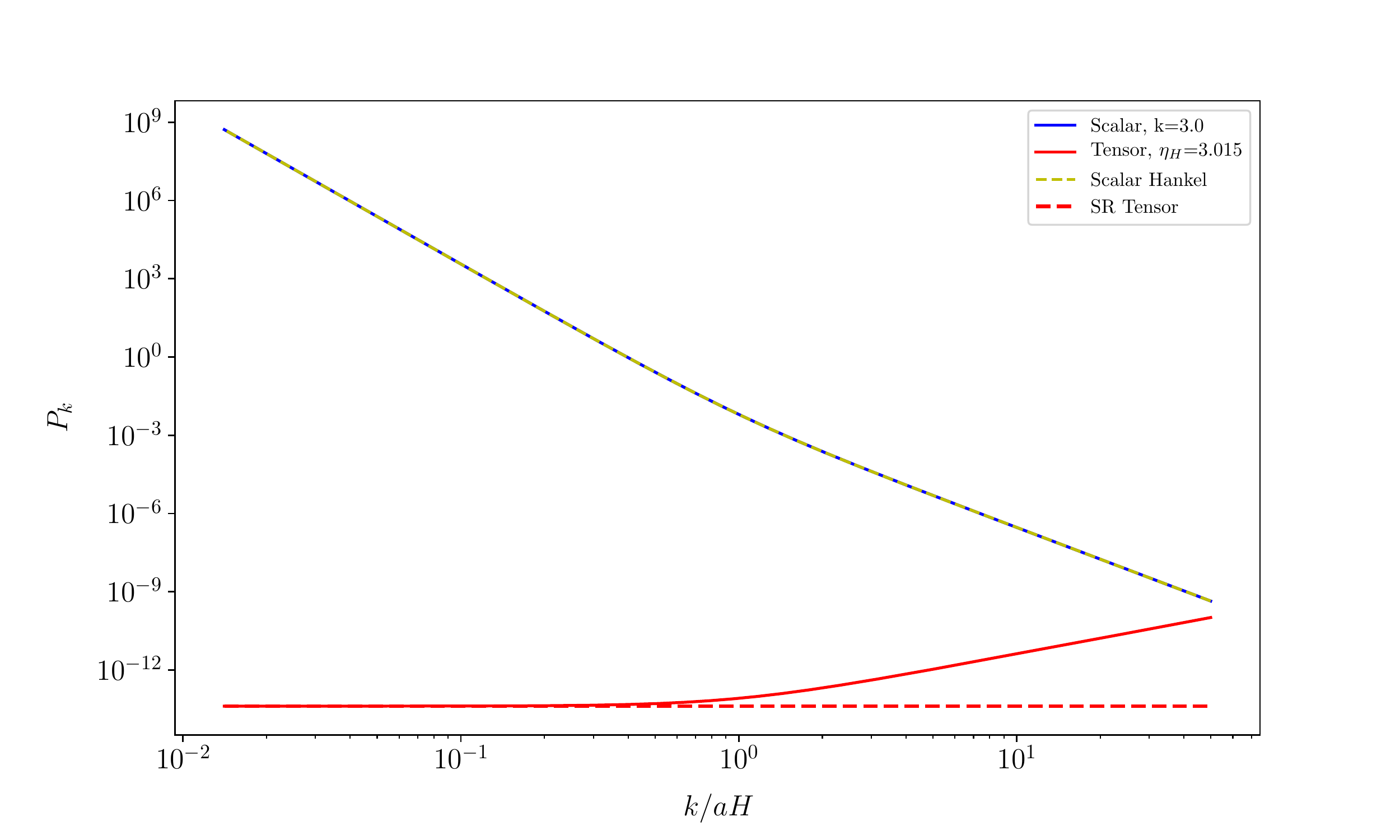}
\caption{The scalar and tensor power spectra for the model \eqref{vsoleq1} with $\eta_H=3.015$. The solid lines are for the numerical results.
The yellow dashed line is the result by using the formulae \eqref{modesol1} with the Hankel function, and the dashed line denotes the asymptotic result with the slow-roll approximation.}
\label{pertfig1}
\end{figure}

\begin{figure}[htp]
\centering
\includegraphics[width=0.6\textwidth]{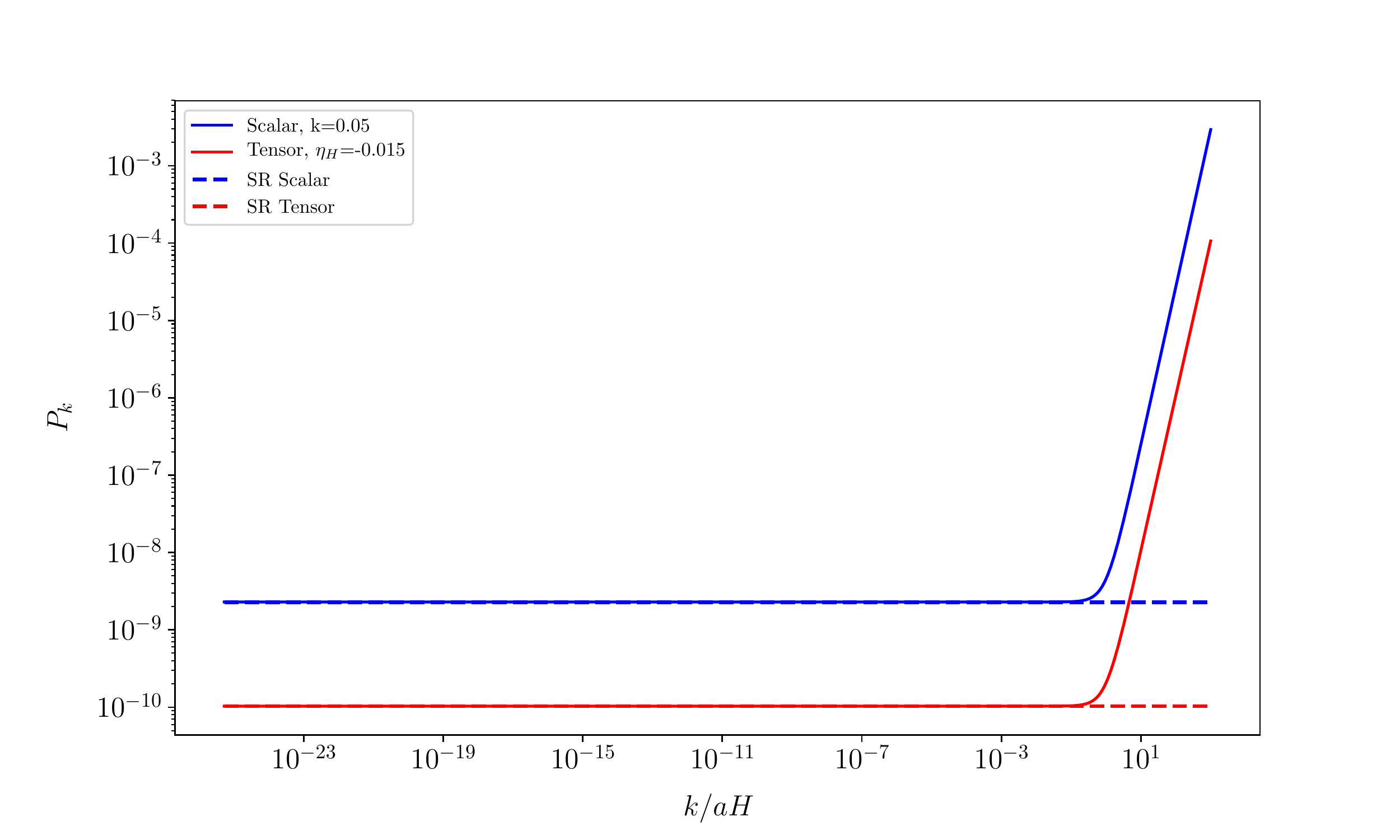}
\caption{The scalar and tensor power spectra for the model \eqref{vsoleq3} with $\eta_H=-0.015$.
The solid lines are for the numerical results,
and the red dashed lines denote the asymptotic results with the slow-roll approximation.}
\label{pertfig2}
\end{figure}

Eqs. \eqref{sraseq1} and \eqref{usraseq1} show that the amplitudes
of the power spectra for both the slow-roll and ultra slow-roll inflation
have the same form.
From Eqs. \eqref{nseqeta} and \eqref{nseqeta1}, we see that the power spectra for both the slow-roll and ultra slow-roll inflation are nearly scale invariant.
If we neglect $\epsilon_H$ in Eqs. \eqref{nseqeta} and \eqref{nseqeta1}, the expressions for the slow-roll inflation with $\eta_H=\alpha$
and the ultra slow-roll inflation with $\eta_H=3-\alpha$ are the same, so it seems that there exists a duality between $\eta_H$ and $\bar{\eta}_H=3-\eta_H$.
In particular, the model \eqref{vsoleq1} is self-dual when $0\le\eta_H\le 3$.
The model \eqref{vsoleq1} with $\eta_H>3$ is dual to the model \eqref{vsoleq3} with $\eta_H<0$.
Note that $\dot{\epsilon}_H<0$ for the model \eqref{vsoleq1}, while $\dot{\epsilon}_H>0$ for the model \eqref{vsoleq3}.
For the model \eqref{vsoleq1} with $\eta_H>3$, the inflaton climbs up instead of rolling down the potential
and the constant-roll inflationary solution is not an attractor \cite{Motohashi:2014ppa}.
Furthermore, as shown in Ref. \cite{Motohashi:2014ppa}, in the model \eqref{vsoleq1} with $\eta_H>3/2$,
the curvature perturbation grows on both the sub-horizon and super-horizon scales,
but the curvature perturbation decreases on the sub-horizon scales and is frozen
on the super-horizon scales in both the model \eqref{vsoleq1} with $\eta_H<1$ and the model \eqref{vsoleq3} as shown in Figs. \ref{pertfig1} and \ref{pertfig2},
so this duality is false because the behaviors of the background and the curvature perturbations
are totally different for the constant-roll inflation with large and small $\eta_H$.
Due to the growth of the curvature perturbations on super-horizon scales for the constant-roll model \eqref{vsoleq1} with $\eta_H>3/2$,
the scalar power spectrum \eqref{pkeq1} should be evaluated at the end of inflation instead of the horizon crossing \cite{Namjoo:2012aa,Cheng:2015oqa,Cheng:2018qof,Byrnes:2018txb,Passaglia:2018ixg}.
For the model \eqref{vsoleq2}, because no inflation happens if $\eta_H>1$, so the duality is inapplicable to this model.
For the same reason, the model \eqref{vsoleq4} is not dual to the model \eqref{vsoleq2}.

Furthermore, $\epsilon_H$ is usually not negligible for the slow-roll inflation while it may be negligible for the ultra slow-roll inflation,
the amplitudes \eqref{sraseq1} and \eqref{usraseq1} for both the scalar and tensor spectra will be different when the effect of $\epsilon_H$ is included,
so there is no duality in the constant-roll inflation with large $\eta_H\approx 3$ and small $\eta_H\approx 0$.
In particular, for the ultra slow-roll inflation, the scalar perturbation may be very large and the tensor to scalar ratio $r$ may be negligible.

\section{The observational constraints}
\label{sec3}

For the slow-roll inflation, in terms of the remaining number of e-folds $N$ before the end of inflation, from Eq. \eqref{dotep}, we get
\begin{equation}
\label{epsn}
\epsilon_H(N)=\frac{\eta_He^{2\eta_HN}}{-1+\eta_H+ e^{2\eta_H N}},
\end{equation}
where we impose the condition of the end of inflation $\epsilon_H(N=0)=1$.
This formulae only applies to the model with $\dot\epsilon_H>0$, like the model \eqref{vsoleq2}.

For the ultra slow-roll inflation, $\epsilon_H$ decreases monotonically with time and inflation does end,
we need some mechanisms to end inflation.
Instead of using $N$, we introduce the
number of $e$-folds $\bar{N}$ after the start of inflation \cite{GalvezGhersi:2018haa}. From Eq. \eqref{dotep}, we get
\begin{equation}
\label{epsna1}
\epsilon_H(\bar{N})=\frac{\eta_H}{1 + e^{2\eta_H (\bar{N}+C)}},
\end{equation}
where $C$ is an integration constant.  Take $N'=\bar{N}+C$, we get
\begin{equation}
\label{epsna2}
\epsilon_H(\bar{N})=\frac{\eta_H}{1+ e^{2\eta_H N'}}.
\end{equation}

\begin{figure}[htbp]
\centering
\includegraphics[width=0.6\textwidth]{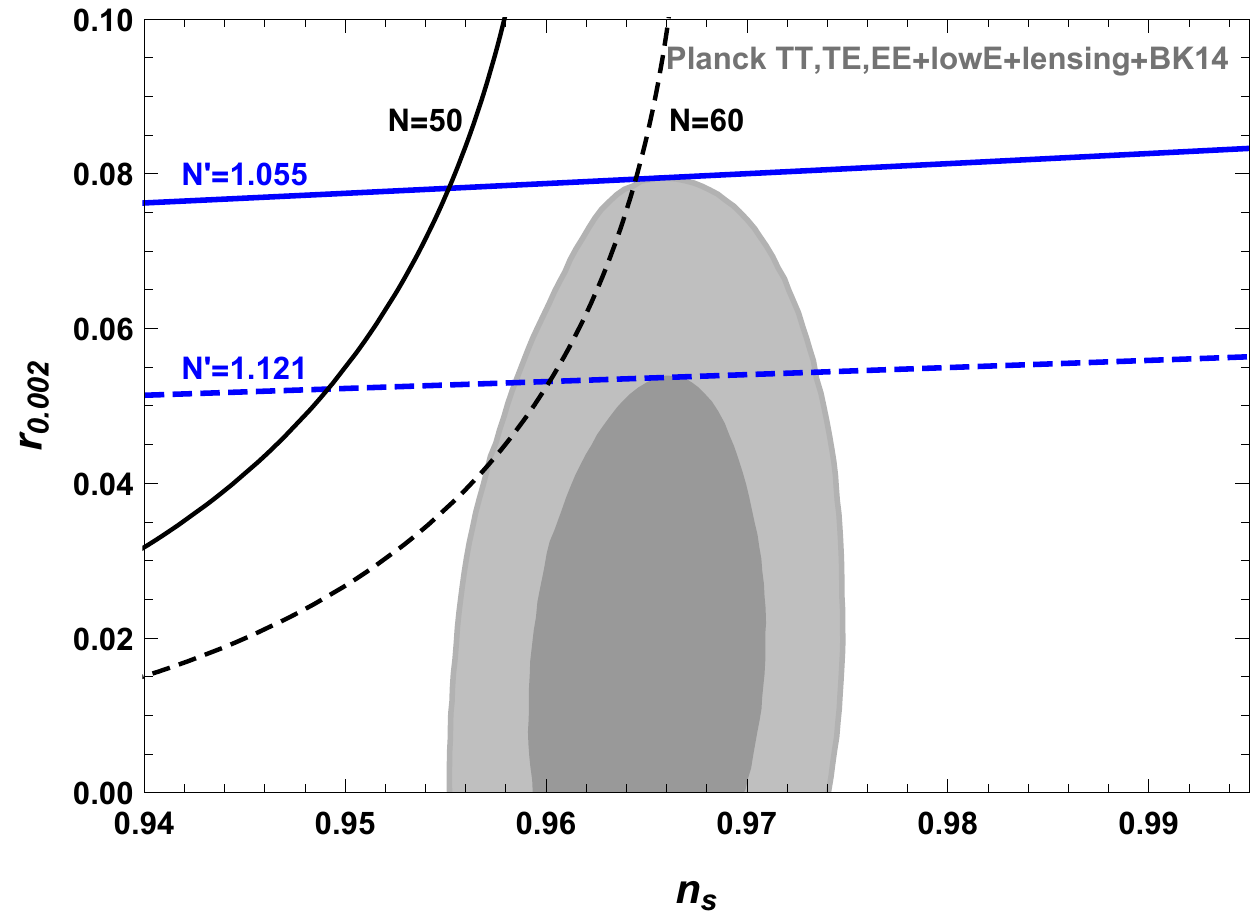}
\caption{The marginalized 68\%, and 95\% confidence level contours for
$n_s$ and $r$ from Planck 2018 and BICEP2 results \cite{Akrami:2018odb,Ade:2018gkx} and the observational
constraints on the constant roll inflationary models.
The black lines represent the model Eq. \eqref{epsn} and the blue lines
denote the model Eq. \eqref{epsna2}.}
\label{ultra}
\end{figure}

\begin{figure}[htbp]
\centering
\includegraphics[width=0.6\textwidth]{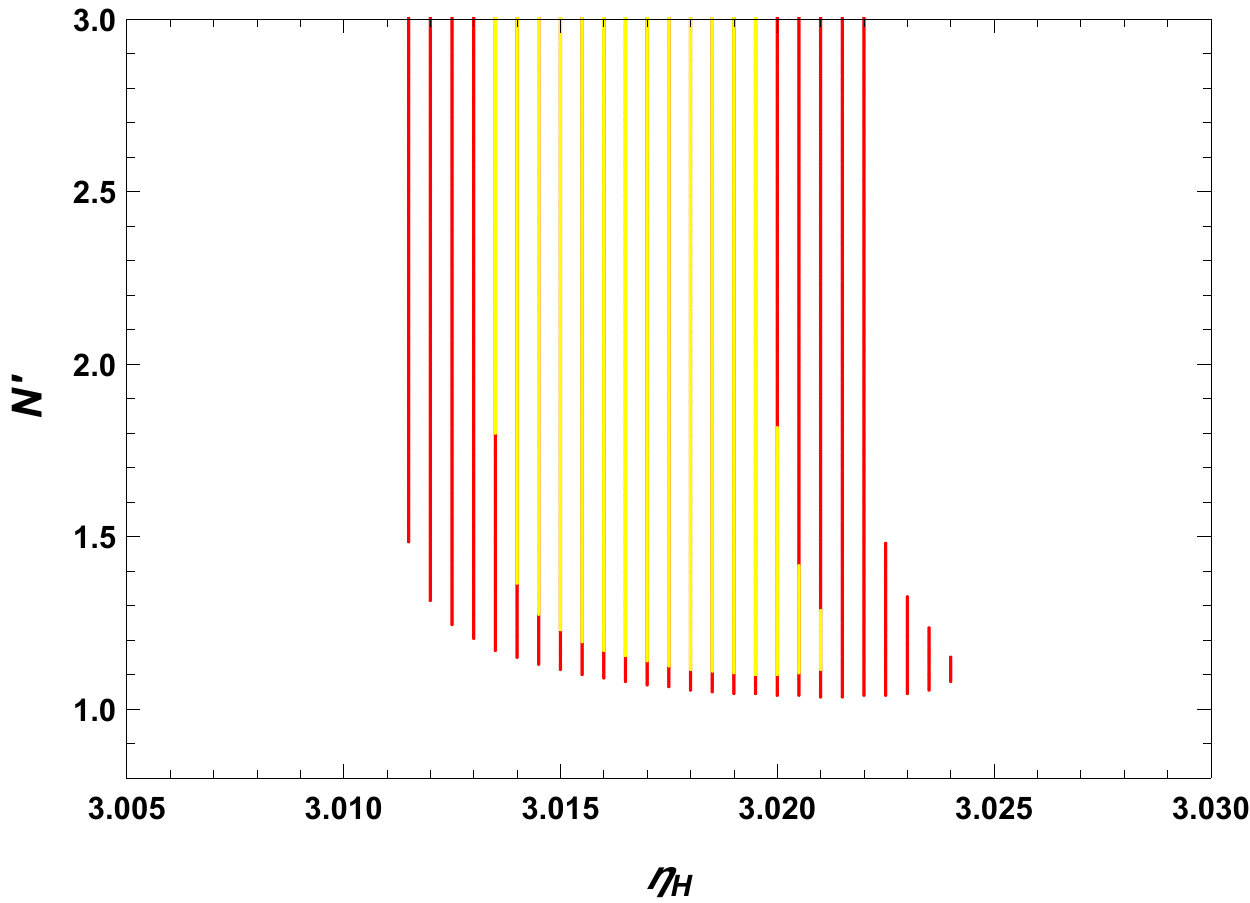}
\caption{The observational constraints on $\eta_H$ and $N'$.
The yellow and red regions correspond to the 68\% and 95\% C.L.s, respectively.}
\label{etaN}
\end{figure}

Substituting Eq. \eqref{epsn} into Eqs. \eqref{nseq1} and \eqref{ra}, we can calculate $n_s$ and $r$
for the constant-roll inflation with increasing $\epsilon_H$.
Substituting Eq. \eqref{epsna2} into Eqs. \eqref{nseq1} and \eqref{ra}, we can calculate $n_s$ and $r$
for the constant-roll inflation with decreasing $\epsilon_H$.
The results along with the Planck 2018 and BICEP2 constraints \cite{Akrami:2018odb,Ade:2018gkx} are shown in Fig. \ref{ultra}.
In Fig. \ref{ultra}, the black lines represent the calculated results with Eq. \eqref{epsn} and the blue lines
denote the calculated results with Eq. \eqref{epsna2}. The model with increasing $\epsilon_H$ is excluded
by the observations if we take $N=50$ (the solid black line)
and is marginally consistent with the observations at the 95\% level if we take $N=60$ (the dashed black line).
The constraint is $-0.016\le\eta_H\le-0.0078$ at the 95\% C.L for $N=60$. The model with decreasing $\epsilon_H$ is consistent with the observations,
we find that $N'\le 1.055$ at the 68\% C.L. and $N'\le 1.121$ at the 95\% C.L.
The constraints on the parameters $\eta_H$ and $N'$ for the model \eqref{epsna2} are shown in Fig. \ref{etaN}.
We get $3.0135\le \eta_H\le 3.021$ at the 68\% C.L., and $3.0115\le \eta_H\le 3.024$ at the 95\% C.L.
These results show that there is no duality between $\eta_H$ and $\bar{\eta}_H=3-\eta_H$.

\section{Conclusions}
\label{sec5}
For the constant roll model \eqref{vsoleq1}, $\dot{\epsilon}_H<0$ and some mechanisms need to be introduced to end the inflation.
The scalar perturbation grows on the sub-horizon scales if $\eta_H>1$ and the super-horizon scales if $\eta_H>3/2$. If $\eta_H>3$,
the inflaton climbs up the potential and the constant roll solution is not an attractor.
For the constant roll model \eqref{vsoleq2}, $\dot{\epsilon}_H>0$ and enough inflation happens only if $\eta_H\ll 1$ because $\epsilon_H>\eta_H$ in this model.

To the first order of $\epsilon_H$, we derive the formulae for $n_s$ and $r$. If we neglect the contribution of $\epsilon_H$
which is a reasonable assumption for the constant roll inflationary models with decreasing $\epsilon_H$ during inflation,
there exists a duality between small $\eta_H=\alpha$ and large $\eta_H=3-\alpha$ with $\alpha\ll 1$ for the expressions of $n_s$ and $r$.
Therefore, it seems that the models \eqref{vsoleq1} and \eqref{vsoleq2} are self dual if $0<\eta_H<3$,
the model \eqref{vsoleq1} with $\eta_H>3$ is dual to the model \eqref{vsoleq3}, the model \eqref{vsoleq2} with $\eta_H>3$ is dual to the model \eqref{vsoleq4}.
As discussed above, there is no inflation in the model \eqref{vsoleq2} with $\eta_H>1$ and
the behaviors of the background and scalar perturbations for the model \eqref{vsoleq1} with $\eta_H>3/2$
are very different from those in the model \eqref{vsoleq1} with $\eta_H<1$ and the model \eqref{vsoleq3},
so the seemingly duality between the constant-roll inflation with large and small $\eta_H$ does not exist.
By fitting the constant roll models to the observations, we find that the model with increasing $\epsilon_H$ is excluded
by the observations if we take $N=50$. If we take $N=60$,
the constraint is $-0.016\le\eta_H\le-0.0078$ at the 95\% C.L.
For the models with decreasing $\epsilon_H$,
we obtain that $3.0135\le \eta_H\le 3.021$ at the 68\% C.L., and $3.0115\le \eta_H\le 3.024$ at the 95\% C.L.
These results confirm that the duality between $\eta_H$ and $\bar{\eta}_H=3-\eta_H$ does not exist.

\begin{acknowledgments}
This research was supported in part by the National Natural Science
Foundation of China under Grant Nos. 11605061 and 11875136,
the Major Program of the National Natural Science Foundation of China under Grant No. 11690021, the Fundamental Research Funds for the Central Universities under Grant Nos. XDJK2017C059 and SWU116053.
Qing Gao acknowledges the financial support from China Scholarship Council for sponsoring her visit to California Institute of Technology,
and thanks California Institute of Technology for the hospitality.
\end{acknowledgments}



%

\end{document}